\documentclass[aps,prd,tighten,11pt,a4,nofootinbib]{revtex4}

\usepackage[utf8]{inputenc}
\usepackage{graphicx}
\usepackage{bm}
\usepackage{amsmath,amssymb}
\usepackage{latexsym}
\usepackage{color}
\usepackage{tabularx}

\newcommand{\mM}{\mathcal{M}}

\newcommand{\bee}{\begin{equation}}
\newcommand{\eee}{\end{equation}}

\begin{document}

\title{Quasinormal modes of dirty black holes in the two-loop renormalizable effective gravity.}

\author{Jerzy Matyjasek}
\email{jurek@kft.umcs.lublin.pl, jirinek@gmail.com}
\affiliation{Institute of Physics,
Maria Curie-Sk\l odowska University\\
pl. Marii Curie-Sk\l odowskiej 1,
20-031 Lublin, Poland}

\begin{abstract}
We consider gravitational quasinormal modes of the static and spherically-symmetric 
dirty black holes in the effective theory of gravity which is renormalizable at the two-loop 
level. It is demonstrated that using the WKB-Pad\'e summation proposed in~\cite{jaOp} one 
can achieve sufficient accuracy to calculate corrections to the complex frequencies 
of the quasinormal modes caused by the Goroff-Sagnotti curvature terms. It is shown that 
the Goroff-Sagnotti correction (with our choice of the sign of the coupling constant) 
increases damping of the fundamental modes (except for the lowest fundamental mode) 
and decreases their frequencies. We argue that the methods adopted in this paper can be 
used in the analysis of the influence of the higher-order curvature terms upon the quasinormal 
modes and in a number of related problems that require high accuracy.
\end{abstract}
%\pacs{04.62.+v,04.70.-s}
\maketitle

\section{Introduction}
The reaction of a black hole to small perturbations is described by a set of oscillations, 
called quasinormal,  and characterized by complex numbers, $\omega,$ the real part of which 
gives the frequency of the mode whereas the imaginary part  controls its damping rate.
Mathematically, the quasinormal modes  considered in this paper are the solutions 
to the ordinary second order Schr\"odinger-like differential equation 
\begin{equation}
\frac{d^{2}}{dx^{2}} \Psi + \left( \omega^{2} - V[r(x)]\right) \Psi = 0
\label{10} 
\end{equation}
with the boundary conditions corresponding to purely outgoing waves at infinity
and purely ingoing waves at the horizon. Here $x$ is the tortoise coordinate, 
$\Psi = \Psi[r(x)]$  describes the radial perturbations in the linear regime and $V$ 
is the potential. The potential is constant as $|x| \to \infty$ and has a maximum at 
$x_{0}.$ For the mode of a given spin weight, $j,$ the quasinormal frequencies are 
labeled by the multipole number, $\ell,$ and the overtone number $n.$

Unfortunately, in the black hole context,
it is impossible to solve this equation exactly and 
consequently one has to resort to numerical and/or approximate methods\footnote{In certain cases the solution 
of~Eq.(\ref{10}) can be expressed in terms of the confluent Heun functions. See, e.g. \cite{fiziev1,fiziev2,hatsuda2} 
and references cited therein. }. 
Since their discovery
by Vishveshwara in 1970~\cite{Vish} an enormous amount of work has been carried out 
on the quasinormal oscillations. Interested readers are referred to the excellent 
reviews~\cite{KokkotasR,NollertR,KonoplyaR,BertiR}
covering almost all aspects of the problem.
Here we mention only the most popular and highly accurate  numerical approaches: 
the method of continued fractions~\cite{Leaver1,nollert2,Rostworowski}, the Hill determinant  
method~\cite{Pancha}, asymptotic iteration~\cite{Nay}, the pseudospectral method~\cite{Jansen} 
and the method of Nollert and Schmidt~\cite{nollert1}. On the other hand, we have a group 
of analytic and semi-analytic methods based on the WKB expansion and its 
variants~\cite{wkb0,wkb1,wkb2,wkb3,wkb4,phase1,phase2,Konoplya6} and the related method 
of Gal'tsov and Matukhin~\cite{Galtsov}. 
Among the WKB-based approximations the most popular are the (third-order) Iyer-Will method~\cite{wkb1}
and its sixth-order generalization constructed by Konoplya~\cite{Konoplya6}.
Moreover, coputationally still very promising is the method developed by 
Zaslavskii~\cite{OlegZ}, who following the ideas of 
Refs.~\cite{blome,val1,val2} reduced the problem to the calculation of the energy levels 
of the quantum anharmonic oscillator. As has been demonstrated  in Ref.~\cite{OlegZ},
one can reproduce the Iyer-Will~\cite{wkb1} results by calculating  the first two nontrivial 
corrections to the energy levels of the sixth-order anharmonic oscillator and this 
equivalence can be extended to higher orders~\cite{jaGo}.

Recently, it has been proposed to construct the Pad\'e
transform of the WKB series describing complex frequencies of the quasinormal modes instead 
of just summing them term by term~\cite{jaOp,jaGo}. This approach  appears to be a major 
improvement over the pure WKB method. Indeed, its has been shown in Refs~\cite{jaOp,jaGo} that 
(within the domain of  applicability)  one can obtain highly accurate values of the quasinormal 
frequencies.  Depending on the number of terms retained in the WKB expansion one can achieve
the accuracy of (at least) 24 decimal places for the low-lying modes\footnote{The WKB 
results have been compared with the results obtained
within the framework of the continued fraction method. The accuracy of the results is  
limited by the available computer resources.}.

The aforementioned techniques have been successfully applied to various black hole systems, too numerous
to list them here. Once again the reader is referred to review papers.  Here we shall discuss
certain aspects of the effective gravity in the context of the quasinormal oscillations of black holes.
The influence of the higher-order curvature terms (see, e.g., Refs~\cite{Myers1,Lu,tel,Dobado} )
on quasinormal modes has 
attracted some attention recently. (See for example Refs.~\cite{Roman_sixth,Hendi,Bouhmadi} 
and the references cited therein). In this paper we shall investigate this problem in some detail. 
We will limit ourselves to the  two-loop renormalizable effective gravity and concentrate on the following issues: 
First, we check if the adapted  
method (which is based on the results of Refs.~\cite{jaOp,jaGo}) is sufficiently sensitive 
to quantify the influence of the higher order terms 
upon the quasinormal modes. Secondly, we compare the complex frequencies calculated for the classical 
black hole and its two-loop counterpart.  Finally, we will briefly discuss the danger 
of relying too much on the schemes that involve only a few first terms of the WKB expansion.

The paper is organized as follows. In Sec.~\ref{sec2} we study the spherically-symmetric black holes
in the two-loop renormalizable effective gravity and give main equations of the problem.
In Sec.~\ref{sec3a} we illustrate the adopted method using simple Mashhoon~\cite{Bahram} 
and Schutz-Will~\cite{wkb0} approach\footnote{Although 
the authors adopted different strategies the resulting equations are essentially the same and 
we will abbreviate them as MSW equations.}
with the Regge-Wheeler and the Zerilli potentials expressed in terms of the Lambert functions. The corrections 
caused by the sixth-order terms are presented graphically.
The accurate calculations of the quasinormal modes are carried out in Sec.~\ref{sec3b}, where we also
study the influence of the second-order corrections to the black hole solution on
the quasinormal frequencies.
Finally, in Sec.~\ref{sec4} we discuss the results obtained and the dangers of naive summation of 
the WKB terms or using simplistic methods.

Throughout the paper we use natural units $c = G =1.$ The signature of the metric is taken to be ``mainly positive'', i.e., $+2,$ and 
the conventions for the curvature tensor are $ \mathcal{R}^{a}_{\, b c d}  = \partial_{c} \Gamma^{a}_{\,b d} ...,$ and 
$\mathcal{R}^{a}_{\, b a c} = \mathcal{R}_{bd}$. 
 
\section{Dirty black holes}
\label{sec2}

As is well-known, the macroscopic black holes are sensitive to the higher-order terms 
in the gravitational action. Typically, such terms are
constructed from the basis of the curvature monomial invariants of definite order and
degree and appear in a natural way in the low-energy limit of
the string theory, phenomenological effecitve Lagrangians  and the Lovelock gravity. 
Moreover, the renormalized one-loop effective action of the quantized
massive fields in the large mass limit is constructed form the curvature invariants
(the type of the field enters through the spin-dependent numerical coefficients).
The  general gravitational action of this type can be written as 
\begin{equation}
S_{g} = \sum_{k=0}^{m} \alpha_{k} S_{k},
\end{equation}
where each $S_{k} $ is constructed from the curvature invariants of the definite order $s$ (the number
of differentiations of the metric) and degree $q$ (the number of factors). Here $s=2k,$ $S_{0}$ is related
to the cosmological term and $S_{1}$ is the  standard Einstein-Hilbert action.
The total action is therefore the  sum of the gravitational action and the matter contribution, where the latter 
may also contain quantum corrections. 
The result of the functional differentiations of the total action with respect to the metric tensor
can generally be written as
\begin{equation}
\mathcal{ R}_{ab} -\frac{1}{2} \mathcal{R} g_{ab} + \Lambda g_{ab} +\mathcal{ P}_{ab} = 8 \pi \left( T_{ab} +  T_{ab}^{(1)} \right),
 \label{eqs}
\end{equation}
where $\mathcal{P}_{ab}$  represents the result of the functional differentiation of the higher-order 
curvature terms,
$T_{ab}$ is the stress-energy tensor of the classical matter, $T_{ab}^{(1)}$ is a small correction  
(presumably of quantum origin) and all the remaining symbols have their usual meaning.  
Both the left and the right hand side of (\ref{eqs}) functionally depends on the metric tensor.
Of course, there is no necessity to introduce $P_{ab}$ and $T_{ab}^{(1)}$ terms simultaneously,
typically we have either one or the other present. It should be noted that when the tensor $T_{ab}^{(1)}$ 
is of purely geometric origin 
it may (with some reservations), equally well, be treated as the object that modifies the left hand side 
of the equations~\cite{jaLamb1,jaLamb2}.

One of the most important and interesting applications of the higher-order theories of gravitation
is the search for their imprints on classical configurations modeled by the solutions 
of the Einstein field equations. This should lead to some definite predictions. Unfortunately, the 
complexity of the problem practically excludes construction of the  exact solutions and one is
forced to adopt either some approximations or refer to numerics. Here we shall choose the first option.
To illustrate the procedure, we consider the simplest case of the spacetime generated by the 
spherically symmetric matter distribution.  The line element describing the spacetime in question can be written as
\begin{equation}
d s^{2} = - e^{-2\psi} \left(1-\frac{2m}{r}\right) d t^{2} 
+ \left(1-\frac{2m}{ r}\right)^{-1} d r^{2} + r^{2} d \Omega^{2},
\label{linel}
\end{equation}
where $m = m(r)$ and $\psi = \psi(r)$ are two functions of the radial coordinate
and $ d \Omega^{2}$ denotes the metric on the unit sphere. The functions $m(r)$ 
and $\psi(r)$ are model-dependent, i.e., they are the solutions of the Einstein
field equations describing the particular model. Now, let us assume that the line
element (\ref{linel}) describes a black hole with the event horizon located at $r =r_{+}.$
In what follows we shall refer to this configuration as `dirty' or `corrected' black hole.
For the line element (\ref{linel}) the field equations~(\ref{eqs}) with the cosmological
constant set to zero assume the form
\begin{equation}
 - \frac{2}{r^{2}}\frac{dm}{dr} + \varepsilon P_{t}^{t} = 8 \pi \left(  T_{t}^{t} + \varepsilon T_{t}^{(1)t} \right)
 \label{e1}
\end{equation}
and
\begin{equation}
  - \frac{2 }{r^{2}}\frac{dm}{dr} -\frac{2 }{r} \frac{d\psi}{dr}\left( 1 -\frac{2m(r)}{r}\right) 
  + \varepsilon P_{r}^{r} =  8 \pi \left(  T_{r}^{r} + \varepsilon T_{r}^{(1)r} \right),
  \label{e2}
\end{equation}
where $\varepsilon$  is the  dimensionless parameter that helps to
keep track of the order of terms in complicated expansions, and as such, it should 
be set to 1 at the end of the calculation. 

 The potential of the gravitational perturbations can be written in the form~\cite{matt1,matt2}
\begin{equation}
V(r)=e^{-2\psi}\left(1-\frac{2m}{r}\right)
\left[\frac{\ell(\ell + 1)}{r^2} - \frac{6 m}{r^3}
+ \frac{2}{r^2} \frac{dm}{dr}+
\frac{1}{r} \left(1-\frac{2m}{r}\right) \frac{d\psi}{dr}\,
 \right].
 \label{poott}
\end{equation}
 With $m(r) =\mathcal{M}$ and 
$\psi(r) =0$ the potential $V(r)$ reduces to the Regge-Wheeler potential of the Schwarzschild black hole.
It belongs to a more general class of potentials  describing scalar, vector 
and gravitational perturbations
\begin{equation}
V(r)=e^{-2\psi}\left(1-\frac{2m}{r}\right)
\left[\frac{\ell(\ell + 1)}{r^2} + (1-j^2) \frac{2m}{r^3}
- (1-j) \, \mathcal{R}_{\theta}^{\theta} \,
 \right],
 \label{potgen1}
 \end{equation}
 where
 \begin{equation}
  j = \begin{cases}
       0 & {\rm for \,scalar} \\
       1 & {\rm for\, vector} \\
       2 & {\rm for \, gravity}
      \end{cases}
 \end{equation}
and the angular components of the Ricci tensor are given by
\begin{equation}
 \mathcal{R}_{\theta}^{\theta} = \mathcal{R}_{\phi}^{\phi} = \frac{2 }{r^{2}} \frac{dm}{dr}+ \frac{1}{r} \left( 1-\frac{2m}{r} \right) \frac{d\psi}{dr}.
\end{equation}

Our discussion has been exact up to this point. Now, let
 us assume that the functions $m(r)$ and $\psi(r)$ have the following expansion
\begin{equation}
 m(r) = \sum_{k=0}^{N} \varepsilon^{k} M_{k}(r) + \mathcal{O}(\varepsilon^{N+1})
 \label{ser1}
 \end{equation}
 and
 \begin{equation}
 \psi(r) = \sum_{k=1}^{N} \varepsilon^{k} \psi_{k}(r)  + \mathcal{O}(\varepsilon^{N+1}),
 \label{ser2}
\end{equation}
where $\varepsilon$ is the dimensionless parameter.
Note that the term $\psi_{0}$ has no independent physical meaning 
and is omitted. The system of the differential equations has to be supplemented 
with the suitable boundary conditions. In what follows we shall relate the additive
integration constant with the total mass of the system measured from infinity $r_{\infty},$ i.e.,
$m(r_{\infty}) = \mathcal{M},$
whereas the second
integration constant can be determined from the natural condition $\psi(r_{\infty}) = 0.$ 
Now, inserting the expansions of the functions $m(r)$ and $\psi(r)$ into Eqs.~(\ref{e1}) and (\ref{e2}) 
and collecting terms with like powers of $\varepsilon,$ one obtains a system of the ordinary differential 
equations of ascending complexity. 

Let us concentrate on pure gravity.
As is well known, the one-loop corrections to the pure classical gravity are quadratic and the divergent terms
calculated by 't Hooft and Veltman have the form~\cite{Tini}
\begin{equation}
 \frac{1}{(4\pi)^{2}(D-4)}\left( \frac{1}{120} \mathcal{R}_{ab}\mathcal{R}^{ab} + \frac{7}{20} \mathcal{R}^{2}\right),
\end{equation}
where $D$  is the dimension. Hence the one-loop divergences of pure gravity vanish on-shell,
the result that can be obtained on the basis of symmetry. 
In their seminal papers, Goroff and Sagnotti~\cite{GoroffA,GoroffB}  showed that at the two-loop level
the divergences of the gravitational action are encoded in the term
\begin{equation}
 \frac{209}{2880 (4 \pi)^{2}(D-4)} \int d^{4} x\,\sqrt{-g}\, \mathcal{R}_{ab}^{\phantom{c}\phantom{d}cd} \mathcal{R}_{cd}^{\phantom{c}\phantom{d}ef} \mathcal{R}_{ef}^{\phantom{c}\phantom{d}ab},
\end{equation}
and thus the Einstein theory of gravitation is not renormalizable. Although this result 
seems to be quite pessimistic, one can think of it  as the indication of possible modifications
of the Einstein gravity. Indeed, 
introducing  the term proportional to
\begin{equation}
 S_{3} = \int d^{4} x\,\sqrt{-g}\, \mathcal{R}_{ab}^{\phantom{c}\phantom{d}cd} \mathcal{R}_{cd}^{\phantom{c}\phantom{d}ef} \mathcal{R}_{ef}^{\phantom{c}\phantom{d}ab}
 \label{six}
\end{equation}
to the total action one obtains, in concord with the philosophy of the effective lagrangians, 
a  simplest generalization of the pure Einstein gravity that  absorbs the divergent term.
At the level of the field equations 
$S_{3}$ introduces the term proportional to
\begin{eqnarray}
\frac{1}{\sqrt{-g}} \frac{\delta}{\delta g_{ab}} S_{3}
&=&
 -12 \mathcal{R}_ {c\phantom{b};d}^{\phantom{c}b\phantom{;\phantom{d}}}\mathcal{R}_    {\phantom{c}\phantom{a}\phantom{;\phantom{d}}}^{ca;d}+12 \mathcal{R}_ {c\phantom{b};d}^{\phantom{c}b\phantom{;\phantom{d}}}\mathcal{R}_    {\phantom{d}\phantom{a}\phantom{;\phantom{c}}}^{da;c}-6 \mathcal{R}_ {cde\phantom{b};i}^{\phantom{c}\phantom{d}\phantom{e}b\phantom{;\phantom{i}}}\mathcal{R}_      {\phantom{c}\phantom{d}\phantom{i}\phantom{a}\phantom{;\phantom{e}}}^{cdia;e}+12\mathcal{R}_ {c\phantom{b};de}^{\phantom{c}b\phantom{;\phantom{d}}\phantom{e}}\mathcal{R}_    {\phantom{c}\phantom{d}\phantom{e}\phantom{a}}^{cdea}
 \nonumber\mathcal{R} \\
 &&
 +12 \mathcal{R}_ {c\phantom{a};de}^{\phantom{c}a\phantom{;\phantom{d}}\phantom{e}}\mathcal{R}_    {\phantom{c}\phantom{d}\phantom{e}\phantom{b}}^{cdeb}
 -12 \mathcal{R}_{cdei}^{\phantom{c}\phantom{d}\phantom{e}\phantom{i}}\mathcal{R}_   {j\phantom{c}\phantom{e}\phantom{b}}^{\phantom{j}ceb}\mathcal{R}_    {\phantom{d}\phantom{j}\phantom{i}\phantom{a}}^{djia}-6 \mathcal{R}_{cd}^{\phantom{c}\phantom{d}}\mathcal{R}_  {ei\phantom{c}\phantom{b}}^{\phantom{e}\phantom{i}cb}\mathcal{R}_    {\phantom{d}\phantom{a}\phantom{e}\phantom{i}}^{daei}+ \frac{1}{2}g_  {\phantom{a}\phantom{b}}^{ab}\mathcal{R}_{cdei}^{\phantom{c}\phantom{d}\phantom{e}\phantom{i}}\mathcal{R}_  {jk\phantom{c}\phantom{d}}^{\phantom{j}\phantom{k}cd}\mathcal{R}_    {\phantom{e}\phantom{i}\phantom{j}\phantom{k}}^{eijk}.\nonumber \\
\end{eqnarray}

Now, let us analyze the influence of the  higher-derivative
terms that may appear in the low-energy effective action functional on the complex
frequencies of the quasinormal modes. To keep the calculations as simple as possible we neglect, 
in concord with our previous discussion,  the four derivative terms and restrict ourselves 
to the first order expansion of the functions $m(r)$ and $\psi(r).$ Additionally we assume that 
the total stress-energy tensor vanishes and the sixth-order term~(\ref{six}) is the only source of the modifications 
of the vacuum field equations. 
The total (effective) action is therefore given by
\begin{equation}
 S_{total} =  \int d^{4}x \sqrt{-g}\, \mathcal{R} - \alpha \int d^{4} x\,\sqrt{-g}\,
 \mathcal{R}_{ab}^{\phantom{c}\phantom{d}cd} \mathcal{R}_{cd}^{\phantom{c}\phantom{d}ef} \mathcal{R}_{ef}^{\phantom{c}\phantom{d}ab}.
\end{equation}
 
Our first task is to solve the field equations.
To this end, let us return to the spherically symmetric line element (\ref{linel}) with (\ref{ser1}) and (\ref{ser2}). 
Now, making a substitution $\alpha\to \varepsilon \alpha$ and subsequently, as has been mentioned earlier, functionally 
differentiating the total gravitational  action with respect to the metric tensor, 
inserting the line element
to the thus obtained system of the differential equations and finally expanding 
the result in the powers of $\varepsilon,$ one obtains a chain of differential equations for $\psi_{i}(r)$ and $M_{i}(r).$
The zeroth-order solution is the Schwarzschild line element characterized by the mass $\mathcal{M},$ 
whereas the first order equations  can be written as
\begin{equation}
 -\frac{2}{r^{2}}\frac{d M_{1}(r)}{dr} + \,{\frac {24  \mathcal{M}^{2} \left( 98\,{ \mathcal{M}}-45\,r 
 \right) }{{r}^{9}}}=0
\label{rown1}
\end{equation}
and 
\begin{equation}
 -\frac{d \psi_{1}(r)}{dr} +\frac{648 \mathcal{M}^{2}}{r^{7}}  =0.
 \label{rown2}
\end{equation}
They can be easily integrated and 
 the perturbative 
solution to the sixth-order gravity field equations is given by
\begin{equation}
 m(r) = \mM  - \alpha \frac{4 \mM^{2}}{r^{6}}\left(49\mM-27 r \right)
 \label{sol1}
\end{equation}
and
\begin{equation}
 \psi_{1}(r) = - \alpha \frac{108 \mM^{2}}{r^{6}},
 \label{sol2}
\end{equation}
where $\varepsilon$ has been put to 1. Since the  black
hole solution is characterized by a total mass as seen by a distant observer, the corrected location of 
the event horizon is 
\begin{equation}
 r_{+} = 2 \mM \left(1+ \frac{5 \alpha}{16 \mM^{4}} \right).
\end{equation}

It should be noted that the solution we just found  is equivalent to the solution
constructed in Ref.~\cite{Dobado}, where  $g_{00}$ and $g_{11}$ have been expanded 
in the powers of  $r^{-1}.$ The coefficients of the expansion satisfy a system of 
algebraic equations. Indeed, inserting  (\ref{sol1}) and (\ref{sol2})
into (\ref{linel}), expanding the result in $\varepsilon$  and finally making
substitution $\alpha \to 16\pi \alpha,$ one obtains precisely the solution presented 
in \cite{Dobado}. We prefer our method simply because it is (in our opinion) more natural
and for higher orders it reduces to simple quadratures. More information is given at the end 
of Sec.~\ref{sec3b}.

Of course, the representation given by (\ref{sol1}) and (\ref{sol2}) is not unique. 
One can, equally well, make use of the another set of conditions: $m(r_{+}) = r_{+}/2$ 
and $\psi(r_{\infty}) = 0,$ where $r_{+}$ is the corrected location of the event horizon.
In what follows, however, we will use the former parametrization and characterize the black hole 
by its total mass as seen by a distant observer rather than the radius of the event horizon.

 \section{Quasinormal modes}
Let us return to our discussion of the quasinormal modes and observe that 
the potential $V(r)$ of the gravitational $(j=2)$ perturbations of the black holes 
described by the line element (\ref{linel}) with  (\ref{sol1}) and (\ref{sol2})  
is given by
\begin{equation}
 V(r) = \frac{1}{r}\left( 1- \frac{2 \mathcal{M}}{r}\right)\left( \frac{L}{r} 
 - \frac{6 \mathcal{M}}{r^{2}}\right) - \frac{8 \mathcal{M}^{2} \alpha}{r^{10}} 
 \left( 528 \mathcal{M}^{2} -549 \mathcal{M} r + 5 \mathcal{M} L r + 135 r^{2} \right),
 \label{lin_pot}
\end{equation}
where  $L = \ell(\ell+1).$  Here we focus on the gravitational modes; the scalar and 
electromagnetic perturbations can be analyzed in a similar manner. It can be easily checked that
(\ref{lin_pot}) vanishes at the event horizon, as expected.
In what follows we also need the radial coordinate of the maximum of the effective potential.
A simple calculation shows that it is given by
\begin{equation}
 r_{0} = r_{(0)} + \alpha\, r_{(1)},
\end{equation}
where
\begin{equation}
 r_{(0)}  =\frac{\left(3 L+9 + \sqrt{9 L^2-42 L+81}\right) \mM}{2 L},
\end{equation}
\begin{equation}
 r_{(1)} = -\frac{4 \mM^2 \left(3 (5 L-549) \mM r_{(0)}+1760 \mM^2+360
   r_{(0)}^2\right)}{r_{(0)}^5 \left(40 \mM^2-4 (L+3) \mM r_{(0)}+L r_{(0)}^2\right)}.
\end{equation}
Asymptotically, as $\ell \to \infty ,$ the leading behavior of $r_{(0)}$ and $r_{(1)}$
is given, respectively, by
\begin{equation}
 r_{(0)} \sim 3 \mM + \frac{\mM}{\ell^2}
\end{equation}
and
\begin{equation}
 r_{(1)} \sim \frac{20}{81 \mM^{3}} + \frac{356}{729 \mM^{3} \ell^{2}}.
\end{equation}
Now, we have all the necessary ingredients to calculate the complex frequencies of the quasinormal modes.

\subsection{The first-order approach}
\label{sec3a}

Our strategy for calculating the quasinormal modes can be illustrated by the following
simple example, that is, nevertheless, valid for $\ell \gg 1.$ It would be instructive 
to analyze it in some detail as the more accurate approaches roughly follow a similar path.
This (first-order) approach is mainly 
due to Mashhoon~\cite{Bahram} and Schutz and Will~\cite{wkb0}, and it leads to the 
following simple and elegant expression 
\begin{equation}
i Q_{0}/\left( Q_{0}''\right)^{1/2} = \left( n+\frac{1}{2} \right),
\label{rel1}
\end{equation}
where $n =0,1,2,..,$ 
$ Q_{0} = \omega^{2} - V_{0}$ and prime denotes differentiation with respect to the 
tortoise
coordinate $x.$ Here, the subscript `0' means that the subscripted quantity has to 
be evaluated at the maximum of the potential. The relation (\ref{rel1}) can be
rewritten in the following simple `ready to use' form
\begin{equation}
 \omega^{2} = V_{0} - i \left(n+\frac{1}{2} \right) \left( 2 Q_{0}''\right)^{1/2}.
 \label{schutz}
\end{equation} 
This formula is the starting point for various more profound analyses and is an indispensable tool
in determining the order of magnitude and the general behaviour of the modes.
Moreover, for more complex potentials (as the one studied here)  the MSW  method allows splitting 
of the quasinormal frequencies into two parts: the classical part and the correction, 
each of which can be calculated and studied independently. It is evident that the methods based 
on the summation of the higher-order WKB terms also share this property. 
Unfortunately, even for such simple approximation as that given 
by (\ref{schutz}), the analytic formulas are too complicated (and not very illuminating) to be shown here. 
Instead, we will present the results of our calculations graphically. 

\begin{figure}
\centering
\includegraphics[width=11cm]{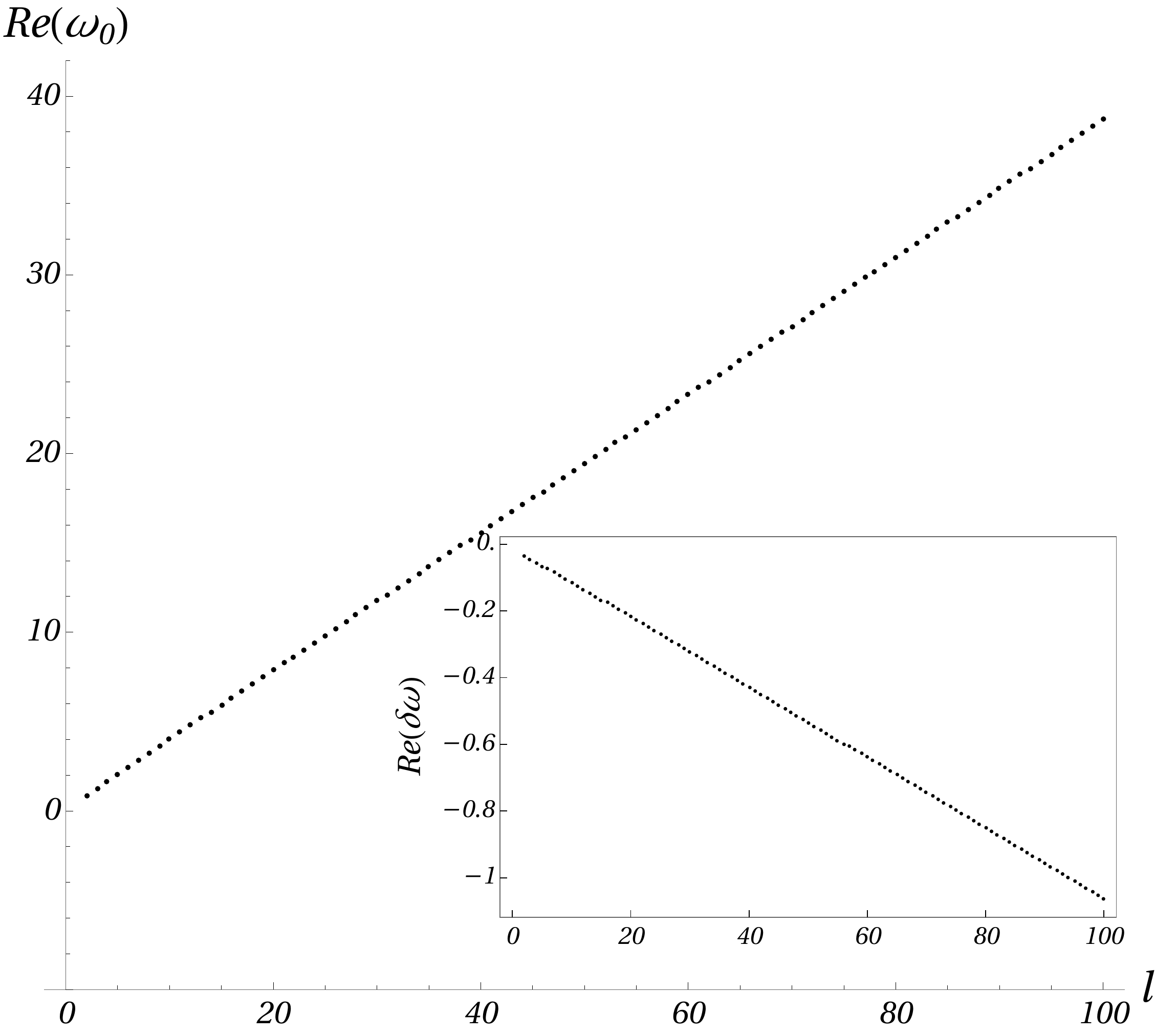}
\caption{The real part of the quasinormal frequencies of the fundamental modes for  $2 \leq \ell \leq 100.$ 
Here $\omega_{0}$ and $\delta \omega$  denote respectively the frequencies of the quasinormal oscillations of 
the classical black hole and their corrections.}
\label{fig1}
\end{figure}

\begin{figure}
\centering
\includegraphics[width=11cm]{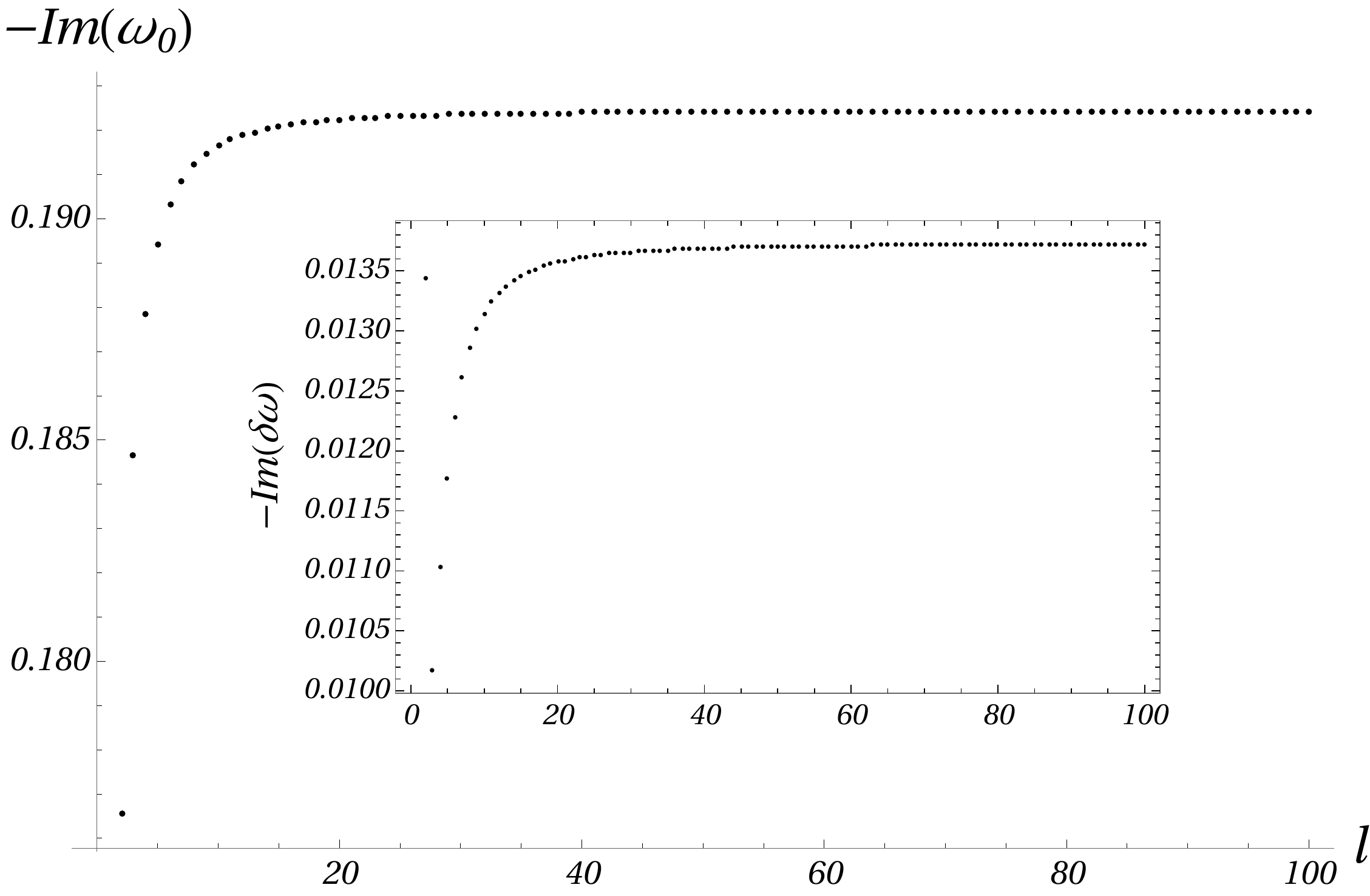}
\caption{The imaginary part of the quasinormal frequencies of the fundamental modes for  $2 \leq \ell \leq 100.$ 
Here $\omega_{0}$ and $\delta \omega$  denote respectively the frequencies of the quasinormal oscillations of 
the classical black hole and their corrections. The asymptotic value of $\Im(\omega_{0})$ and $\Im(\delta \omega)$ as $\ell \to \infty$ is
$-(27)^{-1/2}$ and  $ - \frac{52}{729} (27)^{-1/2},$ respectively.
}
\label{fig2}
\end{figure}

To illustrate the approach we have calculated frequencies of the all fundamental gravitational 
modes for $2 \leq \ell \leq 100.$
The calculated frequencies have the general form
\begin{equation}
 \omega = \omega_{0} + \alpha \delta \omega,
 \label{split}
\end{equation}
where $\omega_{0}$ denotes the frequencies of the classical Schwarzschild black hole, 
$\delta \omega$ is the correction and $\alpha$ is the coupling constant. 
It should be noted that the modifications of the results caused by the 
 two-loop gravity effects are expected to be small and consequently in order to detect them
  very accurate results for both the Schwarzschild and the dirty black hole are needed. 
  Since the formula (\ref{schutz}) gives 
 only qualitative information (although it gets progressively better with increasing $\ell$)
 it cannot be used for the actual comparisons. On the other hand, its simplicity 
 and the fact that for a given potential  both $\omega_{0}$ and $\delta \omega$ are 
 the known (although very complicated) functions  of the parameters $\ell$ and $n$ makes this approach ideal 
 for preliminary analyses. The results of the calculations are plotted in Figs~\ref{fig1} 
 and~\ref{fig2}.
 Inspection of the figures shows that the behavior of the real and 
 the imaginary part of $\delta \omega$ follows the behavior of the Schwarzschild modes.
 Indeed, the linear dependence of $\Re( \omega_{0})$ on $\ell$ is also visible in 
 $\Re(\delta \omega).$ For $\alpha >0,$
 the sixth-order terms tend to decrease the real part of the frequency. Similarly, 
 the imaginary part of the corrections follows the  pattern of $\Im(\omega_{0}),$ 
 making the modes slightly more damped. Finally, observe that both $\Im(\omega_{0})$ 
 and $\Im(\delta \omega)$ asymptotically approach well defined limits. Indeed,
$\Im(\omega_{0}) =-(27)^{-1/2} = -0.192450$  and $\Im(\delta \omega) = - \frac{52}{729} (27)^{-1/2} = -0.013728$ as $\ell \to \infty.$
 
 Let us return to the  Schwarzschild black hole. Inverting  standard relation between the radial and the Regge-Wheeler coordinates
\begin{equation}
 x = r + 2\mathcal{M} \ln \left( \frac{r}{2 \mathcal{M}} -1\right)
\end{equation}
and expressing the result in term of the principal branch of the Lambert $\mathcal{W}$ function\footnote{
The Lambert $\mathcal{W}$ function is defined by the simple relation $\mathcal{W(\xi)}\exp[\mathcal{W(\xi)}] =\xi.$
Other applications of the Lambert functions in the black hole context can be found in Refs.~\cite{jaLamb2,ja_extr,Berej}.
}, 
one has
\begin{equation}
 r = 2 \mathcal{M}\left(1+ \mathcal{W}(y) \right),
\end{equation}
where $y = \exp(x/(2\mathcal{M})-1).$ Now, the Regge-Wheeler potential
can be written in the form 
\begin{equation}
 V_{0} = \frac{\mathcal{W}(y) \left[ \ell (\ell+1) W(y)-3\right]}{4 \left(1+ \mathcal{W}(y) \right)^{4}},
\end{equation}
whereas a slightly more complicated Zerilli potential assumes the form
\begin{equation}
 V_{0} = \mathcal{W}(y) \frac{9+18 \beta \,[1+\mathcal{W}(y)]+ 12 \beta^{2}\,[1+\mathcal{W}(y)]^{2} + 8 \beta^{2}
 (1+\beta)[1+\mathcal{W}(y)]^{3}  }{4 \,[1+\mathcal{W}(y)]^{4} \,[ 3+2 \beta (1+\mathcal{W}(y) ]^{2}},
\end{equation}
where $\beta = (l-1)(l+2)/2.$ 
We prefer this representation over the standard one simply because it depends 
explicitly on the Regge-Wheeler coordinate $x.$
Now, in order to make use of Eq.(\ref{schutz}) it suffices to calculate $x_{0}$ and the second
derivative of the potentials with respect to $x$ at $x_{0}.$ Results for the first 
nine fundamental gravitational modes are tabulated in Table~\ref{tab}.
Even a brief analysis of the results shows that the accuracy is not high. 
Moreover, taking into account  a few additional WKB terms does not necessarily 
improve the quality of the approximation. The foregoing analysis indicates that 
using simple approximation schemes naively, the calculated corrections may be smaller
than the deviations between the approximate and the exact quasinormal frequencies of the 
classical black hole, so care is needed.
\begin{center}
\begin{table}
 \caption{\label{tab} The complex frequencies of the fundamental gravitational quasinormal 
 modes of the Schwarzschild black hole calculated for the Regge-Wheeler potential
 (left column) and  the Zerilli potential (right column). }
\begin{tabular}{|r|c|c|}
\hline\hline
$\ell$ & $\omega_{RW}$  & $\omega_{Z}$ \\ \hline
 2 & $ 0.7976992-0.1765708 i$   & $0.7977882-0.1767022 i$ \\
 3 & $ 1.2331224-0.1846363 i$   & $1.2331234-0.1846375 i$ \\
 4 & $1.6446063-0.1878587 i$    & $1.6446064-0.1878588 i$ \\
 5 & $2.0459245-0.1894270 i$    & $2.0459245-0.1894270 i$ \\
 6 & $2.4420040-0.1903071 i$    & $2.4420040-0.1903071 i$ \\
 7 & $2.8350206-0.1908507 i$    & $2.8350206-0.1908507 i$ \\
 8 & $3.2260873-0.1912102 i$    & $3.2260873-0.1912102 i$ \\
 9 & $3.6158342-0.1914605 i$    & $3.6158342-0.1914605 i$ \\
 10 & $4.0046454-0.1916419 i$   & $4.0046454-0.1916419 i$ \\
\hline\hline
\end{tabular}
\end{table}
\end{center}
A very important lesson that follows from this analysis is the observation that, in principle, 
it should be possible to differentiate between
the `ideal' and the dirty black holes, even if the corrections caused 
by the external factors are small. To do so, however, it is necessary to have a reliable, 
robust and accurate method for calculation the complex frequencies.  Moreover, 
in view of the expected smallness of the corrections the adopted techniques should allow to work with
as many decimal places as needed. 

 \subsection{Pad\'e approximants}
 \label{sec3b}
 
Before we extend the above analysis and make our calculations much more accurate
let us discuss the options we have. First, it would be natural to extend the method of 
calculations along the lines developed by Iyer and Will~\cite{wkb1}. As has been 
demonstrated in Refs.~\cite{wkb1,wkb2,wkb3,Konoplya6,jaOp} it usually gives slightly
more accurate results than its simplified version given by \cite{wkb0}. 
The formula relating the complex frequencies of the quasinormal modes and 
the derivatives of $Q(x)$ at $x=x_{0}$ can be written in the form
\begin{equation}
 \frac{ i Q_{0}}{\sqrt{  2Q''_{0}}\tilde{\varepsilon}} -\sum_{k=2}^{N} 
 \tilde{\varepsilon}^{k-1}\Lambda_{k}= n+ \frac{1}{2},
 \label{master}
\end{equation} 
where the overtones are labeled by $n$ and 
$\tilde{\varepsilon}$ is the expansion parameter that helps to keep track 
of the order of terms in the expansion. The parameter $\tilde{\varepsilon}$ must not be confused with $\varepsilon.$
Each $\Lambda_{k}$ is a combination of the derivatives of $Q(x)$ calculated at $x_{0}$ 
and its complexity grows fast with the order.  The general form of the functions
$\Lambda_{k}$ are known for $k\leq 16$ and, in principle, it is possible  to construct the analog of Eq.(\ref{split}).
However, since the Iyer-Will technique consists of just summing up the $\Lambda$ terms
it cannot be used to obtain highly accurate results. Moreover,  increasing the number of 
$\Lambda$ terms does not improve the quality of the approximation. On the contrary, it can be shown 
that the moduli of the real and imaginary parts of the quasinormal frequencies rapidly grow with the number
of the terms of  WKB series summed. 

A second approach, and the one that will be used here, consists of treating the right hand side of the expression
\begin{equation}
 \omega^{2} = V(x_{0})-i \left(n+\frac{1}{2} \right)  \sqrt{  2Q''_{0}} \tilde{\varepsilon} -i 
 \sqrt{  2Q''_{0}} \sum_{i=2}^{N} \tilde{\varepsilon}^{j} \Lambda_{j}
 \equiv V(x_{0})  + \sum_{i=1}^{N} \tilde{\varepsilon}^{i} \tilde{\Lambda}_{i}
 \label{omm}
\end{equation}
as the power series and instead of summing the terms (which is a bad strategy) we construct 
the Pad\'e approximants~\cite{jaOp,jaGo}. The Pad\'e approximants 
of a formal power series $\sum a_{k} \tilde{\varepsilon}^{k}$ are defined 
as the unique rational functions $\mathcal{P}_{N}^{M}(\tilde{\varepsilon})$ 
of degree $N$ in the denominator and $M$ in the numerator 
satisfying~\cite{CarlB}
\begin{equation}
 \mathcal{P}_{N}^{M}(\tilde{\varepsilon}) - \sum_{k=0}^{M+N} a_{k}\tilde{\varepsilon}^{k}
 ={ \cal{O}}(\tilde{\varepsilon}^{M+N+1}).
\end{equation}
It has been shown that this simple strategy yields amazingly accurate results. For  example,
it can be demonstrated that for the low-lying fundamental gravitational modes of the Schwarzschild 
black hole one  can easily  achieve  accuracy of 32 decimal places or  
better. Such accuracy is a must as we  are interested in the corrections to $\omega_{0}$
caused by the very subtle effects.
The Pad\'e summation of the WKB terms in Eq.(\ref{omm}) has been introduced in Ref.~\cite{jaOp} and subsequently 
extended in Ref.\cite{jaGo} to which the interested reader is referred  for the technical 
details and a general discussion. Although the functions $\Lambda_{k}$
for $k \geq 17$ are unknown, they can be constructed for a given potential with prescribed 
$\ell$ and $n$ numerically~\cite{jaGo,hatsuda,sulejman}. Since the approach is numerical it is practically impossible to
construct the complex frequencies of the quasinormal modes for a general coupling constant.
On the other hand, the calculations can be repeated as many times as needed with various choices
of the coupling constant $\alpha,$ and, consequently, given the expected benefits, 
the loss of the analyticity  in the coupling constant can be treated as a minor sacrifice.
\begin{figure}
\centering
\includegraphics[width=12cm]{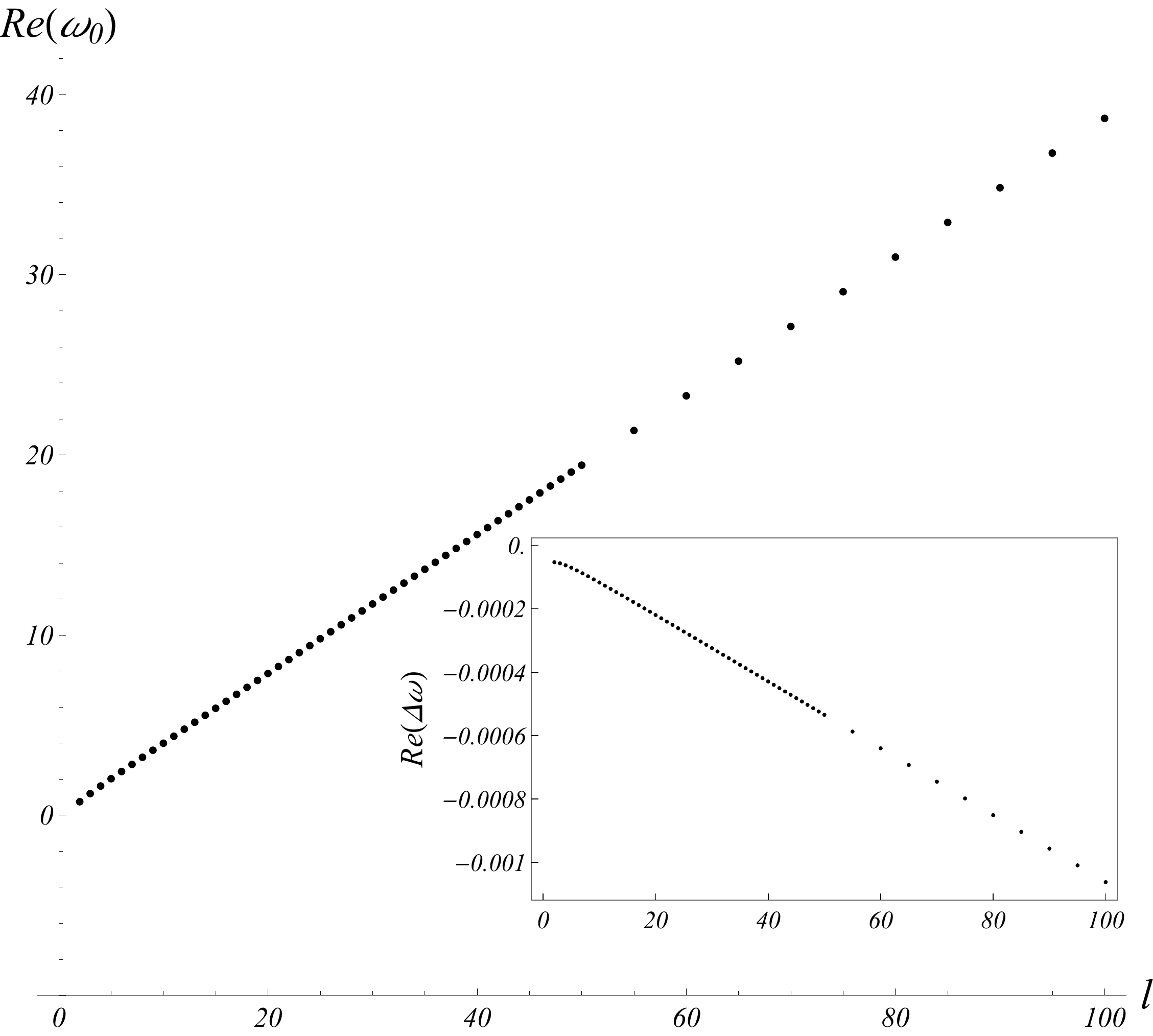}
\caption{The real part of the quasinormal frequencies of the fundamental modes for  $2 \leq \ell \leq 100$ calculated for 
$\alpha = 10^{-3}.$
Here $\omega_{0}$ and $\Delta \omega$  denote respectively the frequencies of the quasinormal oscillations of 
the classical black hole and their corrections. As the quality of the approximation grows with $\ell,$ starting with $\ell =50$ 
we  reduced the number of 
calculated modes. }
\label{fig3}
\end{figure}
\begin{figure}
\centering
\includegraphics[width=12cm]{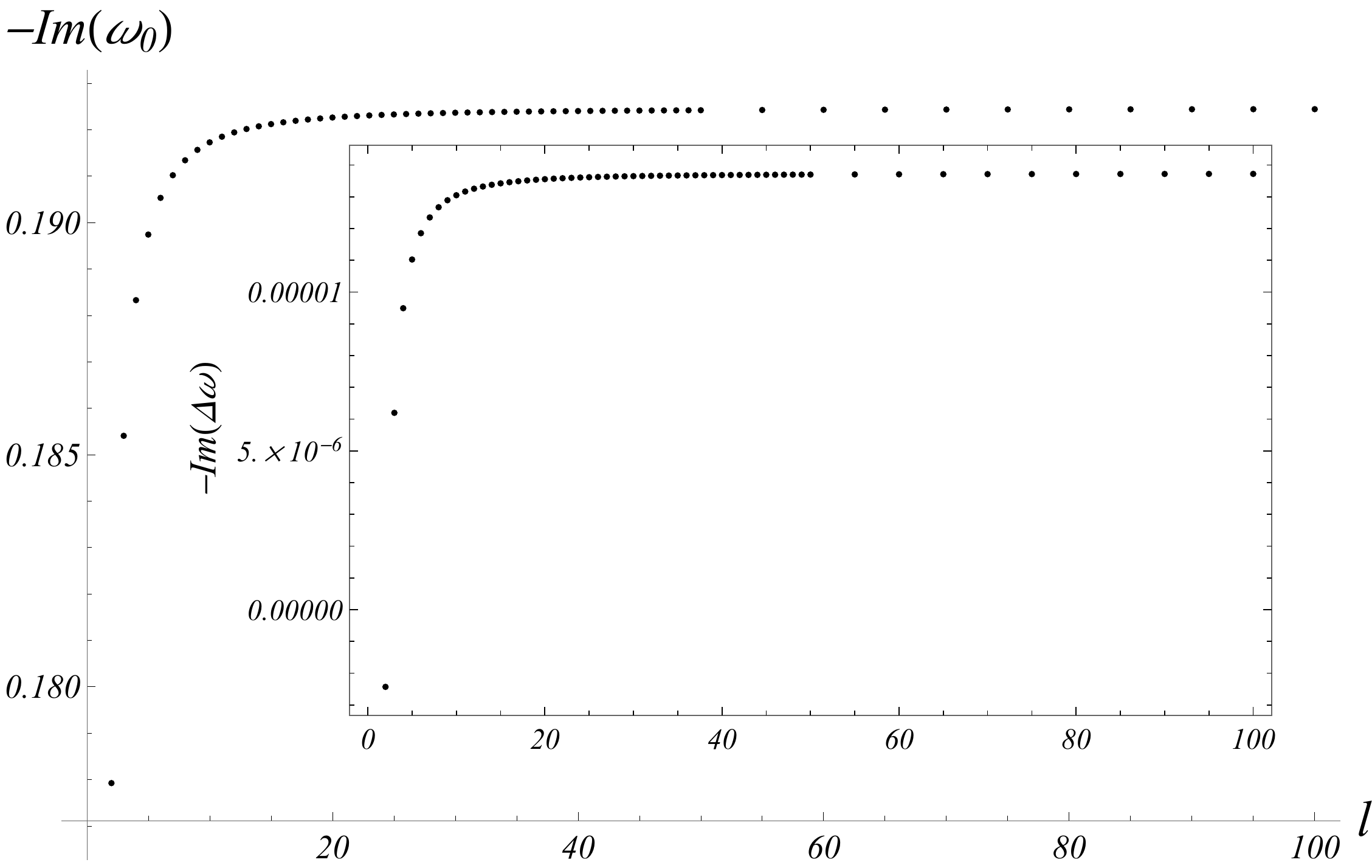}
\caption{ The imaginary part of the quasinormal frequencies of the fundamental modes for  $2 \leq \ell \leq 100$ calculated for 
$\alpha = 10^{-3}.$ 
Here $\omega_{0}$ and $\Delta \omega$  denote respectively the frequencies of the quasinormal oscillations of 
the classical black hole and their corrections. As the quality of the approximation grows with $\ell,$ starting with $\ell =50$
we  reduced the number of 
calculated modes. }
\label{fig4}
\end{figure}
 
 Since we do not know  the coupling parameter $\alpha$ and the adopted 
 method requires knowledge of its numerical value, we shall consider a toy model in which 
 $\alpha = 10^{-3}.$ Such a choice, although unphysical, guarantees  that the corrections 
 will be easily visible in the final results. Of course the method is capable of a 
 very high precision and allows for much smaller values of $\alpha$ as will be demonstrated explicitly 
 at the end of this section. 
 
 Because of the nature of the problem at hand
 we want to (numerically) construct the quantities $\omega$ and $\Delta \omega(\alpha) $
 that satisfy 
 \begin{equation}
  \omega = \omega_{0} + \Delta \omega(\alpha) 
 \end{equation}
 and $\Delta \omega(\alpha) \to 0$ as $\alpha \to 0.$ 
Before we start the presentation of the results let us briefly discuss the general features of the method.
First, it should be observed that for a given $N,$  the accuracy of the Pad\'e transform $\mathcal{P}_{N}^{N}$
increases with $\ell$ and decreases with $n.$ On the other hand, increasing of $N$  improves the accuracy
of the overtones. Of course, for each problem there is a minimal $N$ starting with which one obtains sensible
results.
Since we are interested in a  moderate accuracy of the fundamental gravitational quasinormal modes,
say up tp 20 decimal places, it suffices to start with $\mathcal{P}_{150}^{150}$  and gradually 
decrease $N$ with  increasing  $\ell.$ 
For example, it suffices to take $N=40$ for $\ell = 10.$ Unfortunately, for more complex potentials this places 
severe demands on the computer resources.
%%%%%% Table has been corrected (15 July 2020)
\begin{center}
\begin{table}
 \caption{\label{tabA} The complex frequencies of the fundamental gravitational quasinormal modes of the 
 Schwarzschild black hole
 (left column) and  the dirty black hole (right column) calculated for $\alpha =10^{-3}.$ The  Pad\'e 
 approximants of the WKB series, $\mathcal{P}_{N}^{N},$ are calculated for $N=150.$ }
\begin{tabular}{|r|c|c|}
\hline\hline
$\ell$ & $\omega_{0}$  & $\omega$ \\ \hline
 2 & $0.747343368836083672-0.177924631377871397 i$  & $ 0.747289996857394327-0.177922202116315106 i$ \\
 3 & $1.198886576874980146-0.185406095889895208 i$  & $ 1.198829745091059250-0.185412294358677386 i$ \\
 4 & $1.618356755064478281-0.188327921977846499 i$  & $ 1.618293764592806149-0.188337413836831943 i$ \\
 5 & $2.024590624270701002-0.189741032163219024 i$  & $ 2.024519942903865805-0.189752057406537987 i$ \\
 6 & $2.424019641304260981-0.190531691684164105 i$  & $ 2.423940385532620070-0.190543545087114363 i$ \\
 7 & $2.819470241218645086-0.191019258552094436 i$  & $ 2.819381873352072298-0.191031608642424504 i$ \\
 8 & $3.212387456545430050-0.191341402053726517 i$  & $ 3.212289629226929000-0.191354073479312716 i$ \\
 9 & $3.603589562167645343-0.191565498650669442 i$  & $ 3.603482039130982745-0.191578390083493717 i$ \\
10 & $3.993575588236105108-0.191727744259109785 i$  & $ 3.993458201820883910-0.191740793053382124 i$ \\
\hline\hline
\end{tabular}
\end{table}
\end{center}
Now, in order to obtain $\Delta\omega$  for a given $\ell$ we calculate both $\omega_{0}$ and $\omega.$
Since the quality of the approximation grows with $\ell,$ starting with $\ell =50$ we  reduce the number of 
calculated modes. Inspection of Figs.~\ref{fig3} and ~\ref{fig4} shows that the quasinormal frequencies 
calculated within the framework of the Pad\'e-WKB approach and the MSW method  follow a similar pattern. 
Of course, the latter method is unable to provide required accuracy of the calculations.  Once again we see 
that for a given $\ell$ and a positive $\alpha$ the quasinormal modes of the corrected black hole are more 
suppressed (except for the lowest fundamental mode) whereas their frequency is decreased.  
The results of the calculations (rounded to 19 decimal places) 
are presented in Table~\ref{tabA}. We believe that they are correct to the assumed accuracy. 

\begin{figure}
\centering
\includegraphics[width=11cm]{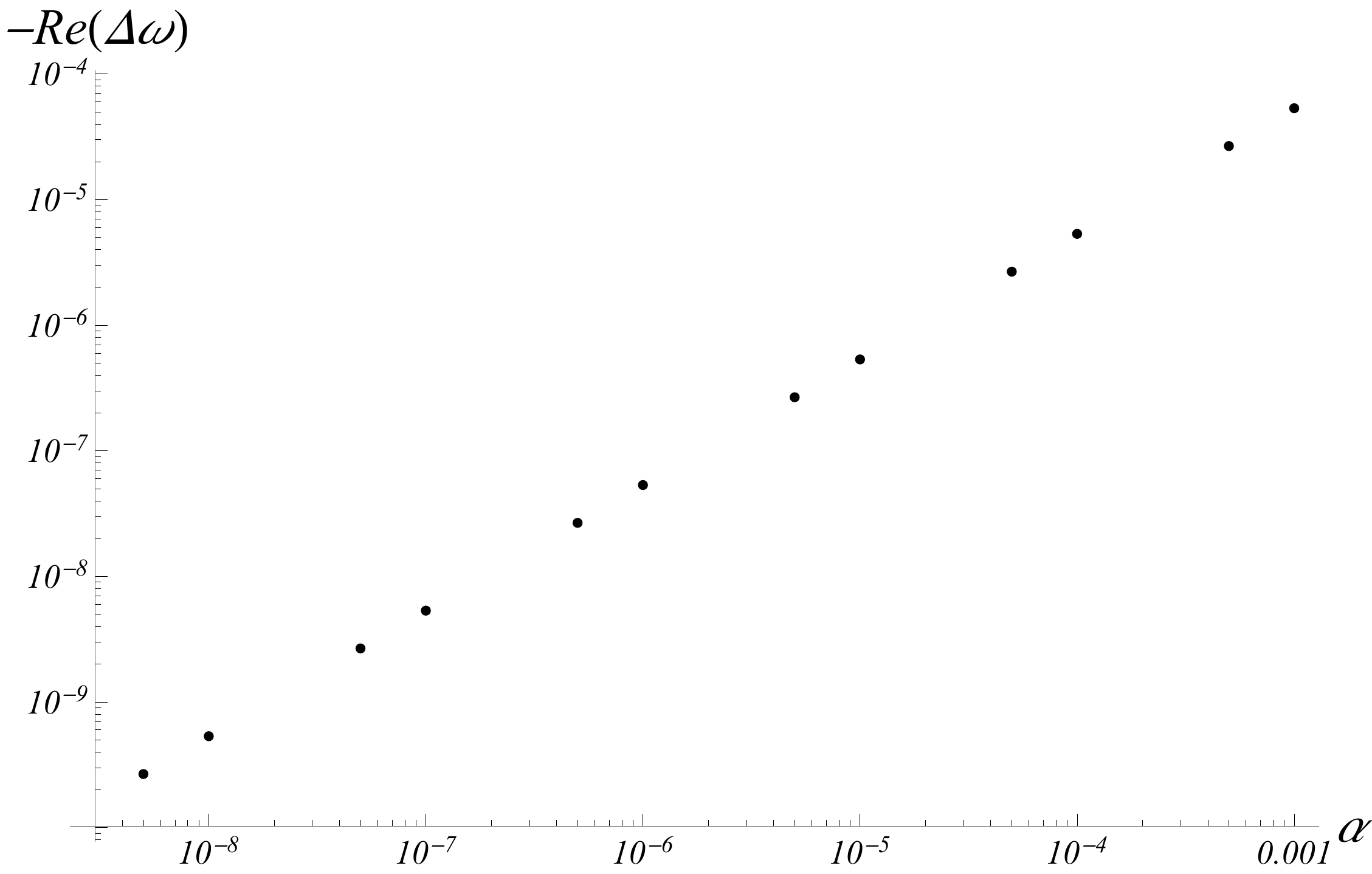}
\caption{The log-log plot of the $- \Re(\Delta\omega )$ part of the gravitational fundamental 
mode $(\ell=2, n=0)$ for  $\alpha  = 5\times 10^{-9}, 10^{-8}, 5 \times 10^{-8}, 
10^{-7}, 5 \times 10^{-7},\dots,10^{-3}$ }
\label{fig5}
\end{figure}

\begin{figure}
\centering
\includegraphics[width=11cm]{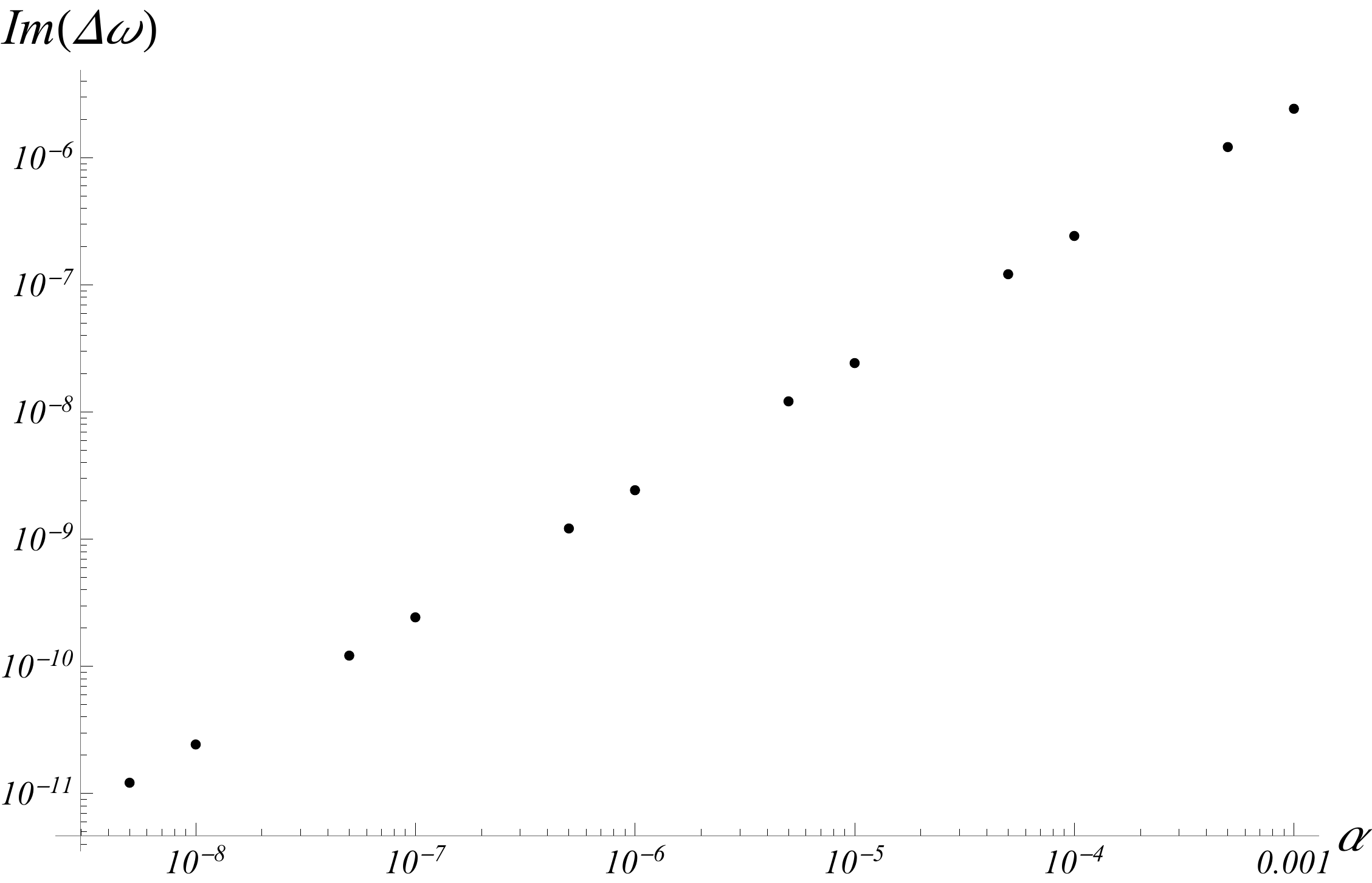}
\caption{ The log-log plot of the $\Im(\Delta\omega )$ part of the gravitational 
fundamental mode $(\ell=2, n=0)$ for $\alpha = 5 \times 10^{-9}, 10^{-8}, 
5 \times 10^{-8}, 10^{-7}, 5 \times 10^{-7},\dots,10^{-3} $ }
\label{fig6}
\end{figure}

It should be noted that with the increase of $\ell$  the stabilization of results is achieved for lower values 
$N$ in $\mathcal{P}_{N}^{N}$ and this observation may speed up the calculations considerably. 
Moreover, inspection of Table~\ref{tabA} and Figs.~\ref{fig5} and~\ref{fig6}  
shows that even with such moderate accuracy it is possible 
to detect the influence of the Goroff-Sagnotti term for $\alpha$ of order $10^{-14}.$ 
In the log-log plots (Figs.~\ref{fig5} and~\ref{fig6})
both $-\Re(\Delta \omega)$ and $\Im(\Delta \omega)$ of the gravitational fundamental mode $(\ell =2, n=0)$ 
calculated for $\alpha = 10^{-8}, 5 \times 10^{-8}, 10^{-7}, 5 \times 10^{-7},\dots,10^{-3} $ lie
on a straight line, an expected result which, nevertheless, can be regarded as the useful check 
of the correctness of the  calculations. For $\ell >2$  the corrections 
follow the same pattern for $-\Re(\Delta \omega)$ and $-\Im(\Delta \omega).$


\subsection{The second-order solution }

Finally, let us consider the influence of the second-order solution of the equations 
(\ref{e1}) and (\ref{e2})  upon the quasinormal modes. Now, for $m(r)$ and $\psi(r)$ one has
\begin{equation}
  m(r) = \mM  - \alpha \frac{4 \mM^{2}}{r^{6}}\left(49\mM-27 r \right) 
  + \frac{24 \alpha ^2 \mathcal{M}^4 (6787 \mathcal{ M} -4104 r)}{11 r^{12}}
  \label{m2}
\end{equation}
and
\begin{equation}
 \psi(r)  = - \alpha \frac{108 \mM^{2}}{r^{6}} +  \frac{1296 \alpha ^2\mathcal{ M}^3 (253 \mathcal{M}-128 r)}{11 r^{12}},
 \label{ps2}
\end{equation}
whereas the event horizon $r_{+}$ is located at
\begin{equation}
 r_{+} = 2\mM \left(1 +  \frac{5 \alpha}{16 \mM^{4}}  -   \frac{1623 \alpha^{2} }{5632 \mM^{8}} \right).
\end{equation}
Making use of Eq.~(\ref{poott}) 
one obtains the following expression describing the effective potential
\begin{eqnarray}
 V(r) &=& \frac{1}{r}\left( 1- \frac{2 \mathcal{M}}{r}\right)\left( \frac{L}{r} 
 - \frac{6 \mathcal{M}}{r^{2}}\right) \nonumber \\
 &&- \frac{8 \mathcal{M}^{2} \alpha}{r^{10}} 
 \left( 528 \mathcal{M}^{2} -549 \mathcal{M} r + 5 \mathcal{M} L r + 135 r^{2} \right)\nonumber \\
 && -\frac{48 \alpha ^2 \mM^3 }{11 r^{16}}\left(-29249 L \mM^2 r+28728 L \mM r^2-6912 L r^3
 \right. \nonumber \\
&&\left. +436656 M^3-567867 \mM^2 r+250128 \mM r^2-38016 r^3\right).
\end{eqnarray}
Repeating the steps of Sec.~\ref{sec3b} one can construct the quasinormal frequencies.
The results of our calculations  are tabulated in Tab.~\ref{tabC}. This time 
however, the calculations are more complex and time consuming as the construction of the derivatives 
of $V(x)$ could impose severe demands on the computer resources.

\begin{center}
\begin{table}
 \caption{\label{tabC} The complex frequencies of the fundamental gravitational quasinormal modes of the 
 dirty black hole with the second-order corrections calculated within the framework of the WKB-Pad\'e technique. 
 The geometry of the black hole is characterized by Eqs.~(\ref{m2}) and (\ref{ps2}). }
\begin{tabular}{|r|c||}
\hline\hline
$\ell$ &  $\omega$ \\ \hline
 2  & $0.747290083278501734-0.177922209944631719 i $\\
 3  & $1.198829835396153738-0.185412278758324958 i $\\
 4  & $1.618293857305014600-0.188337388043360077 i $\\
 5  & $2.024520040450694490-0.189752027161955573 i $\\
 6  & $2.423940490052797489-0.190543512760709371 i $\\
 7  & $2.819381986313369047-0.191031575267004615 i $\\
 8  & $3.212289751605332245-0.191354039537637231 i $\\
 9  & $3.603482171586300389-0.191578355817096047 i $\\
 10 & $3.993458344812290014-0.191740758590793688 i$ \\
\hline\hline
\end{tabular}
\end{table}
\end{center}
Inspection of tables~\ref{tabA}  and~\ref{tabC} shows that the absolute value of the difference
between the first and the second-order results is a few orders of 
magnitude smaller than the difference between  the Schwarzschild 
and the first order results. Indeed, in the first case the difference of the real
part does not exceed $1.5 \times 10^{-7}$ whereas the imaginary part is always smaller than
$3.5 \times 10^{-8}.$ This can be contrasted with the first case, where the analogous differences
are typically $10^3$ times bigger. 

\section{Final Remarks}
\label{sec4}

In this paper, we have investigated the influence of the effective two-loops renormalizable gravity 
upon the quasinormal modes. The idealized ``experimental'' situation we have in mind is the following: 
we have two black holes characterized by the same mass $\mathcal{M}.$ One of them is described by 
the Schwarzschild line element whereas the second one has (presumable small) corrections caused 
by the Goroff-Sagnotti sixth-order curvature terms. Our task is to determine which black hole is which.
We see that this question - when addressed naively - may lead to incorrect answer. Indeed, making use 
of unsophisticated claculational techniques one can obtain results in which the error of the method 
is bigger than the expected effect, so the results, although mathematically correct, do not reflect the
actual situation. 
For example, $\Re(\omega)$ of the lowest fundamental mode of the dirty black hole clculated within 
the framework of the MSW method is closer to the exact Schwarzschild value than
its uncorrected counterpart.
Assuming that the coupling constant $\alpha$ is small all we need is a very accurate 
and sensitive method for calculations of the complex frequencies. In this paper we argue that  
the Pad\'e approximants of the (formal) WKB series describing quasinormal frequencies of the black holes
may have desired features. In the case in hand, one can easily approach the accuracy of, say, 30 decimal places
(or more)
even for low-lying modes.
For example, both the continued fraction method and the WKB-Pad\'e summation agree that to 32 digits accuracy
\begin{equation}
 \omega = 0.74734336883608367158698400595410 -
 0.17792463137787139656092185436905 i
\end{equation}
for the lowest fundamental gravitational mode of the Schwarzschild black hole.
Of course, the method have some limitations, but because of its simplicity 
we believe that it can be the method of choice in many calculations of this type. We have limited ourselves to 
the two-loop renormalizable effective gravity. It is clear that this approach is easily adaptable to other theories
(not necessarily pure gravity) with the higher-order curvature terms and in many related problems.

Finally, a few words on the computational side of the problem are in order. The calculations can be 
roughly divided into the three parts. First, we 
calculate the derivatives of the potential with respect to the $x$ coordinate at $x_{0}.$ Although highly algorithmic, 
this stage (when performed analytically) can be both time and memory consuming. Subsequently  we construct the $\tilde{\Lambda}$ 
functions and finally we calculate the Pad\'e 
transforms of the WKB series. It should be noted that the time spent on calculations of the Pad\'e transforms 
is only a small fraction of the total time of computations. On the other hand, for a given $N,$ the 
calculation time of the WKB series is practically insensitive to the type of the black hole. All the calculations
presented in this paper can easily be completed on a budget laptop  with  16 GB of RAM.

%\bibliography{wkb}

\end{document}